# Percolative nature of the dc paraconductivity in the cuprate superconductors

Petar Popčević,[1,2,†] Damjan Pelc,[3,†,#] Yang Tang,[4] Kristijan Velebit,[1] Zachary Anderson,[4] Vikram Nagarajan,[4] Guichuan Yu,[4] Miroslav Požek,[3*] Neven Barišić[1,3,4*] & Martin Greven[4*]

[1]Institute of Solid State Physics, TU Wien, 1040 Vienna, Austria

[2]Institute of Physics, Bijenička cesta 46, HR-10000, Zagreb, Croatia

[3]Department of Physics, Faculty of Science, University of Zagreb, Bijenička cesta 32, HR-10000, Zagreb, Croatia

[4]School of Physics and Astronomy, University of Minnesota, Minneapolis, MN 55455, USA

[†]These authors have contributed equally

[#]present address: School of Physics and Astronomy, University of Minnesota, Minneapolis, MN 55455, USA

*Correspondence to: mpozek@phy.hr, neven.barisic@tuwien.ac.at, greven@umn.edu

**We present an investigation of the planar direct-current (dc) paraconductivity of the model cuprate material $HgBa_2CuO_{4+\delta}$ in the underdoped part of the phase diagram. The simple quadratic temperature-dependence of the Fermi-liquid normal-state resistivity enables us to extract the paraconductivity above the macroscopic $T_c$ with great accuracy. The paraconductivity exhibits unusual exponential temperature dependence, with a characteristic temperature scale that is distinct from $T_c$. In the entire temperature range where it is discernable, the paraconductivity is quantitatively explained by a simple superconducting percolation model, which implies that underlying gap disorder dominates the emergence of superconductivity.**

The nature of the metallic normal state and of the emergence of superconductivity in the cuprates belong to the most extensively debated problems in condensed matter physics [1]. At temperatures above the macroscopic superconducting transition temperature $T_c$, there exists no long-range coherence, yet traces of superconductivity remain observable, and different experimental investigations have led to widely disparate conclusions [2-14]. In contrast to prevailing thought, it was recently proposed that the normal state of underdoped cuprates exhibits Fermi-liquid charge transport [15-18], and that superconductivity emerges from this state in a percolative manner [7]. Direct-current (dc) conductivity is a highly sensitive probe that can, in principle, provide a unique opportunity to test the consistency of these ideas. Furthermore, the effective-medium approximation required to model such a mixed regime is well established for the dc conductivity response [19,20], whereas calculations for other observables (e.g., magnetic susceptibility) are very challenging. In this Letter, we present benchmark dc conductivity data for a pristine cuprate compound along with modeling results that support both the Fermi-liquid nature of the normal state and the percolative superconductivity emergence in a quantitative manner.

The principal problem in previous investigations of the pre-pairing regime in cuprates has been the separation of the superconducting response from the normal-state response. Different experimental probes can be sensitive to distinct aspects of the normal state. Moreover, it is well established that the underdoped cuprates also exhibit other electronic ordering tendencies including charge-density-wave order [21-27], which has further precluded an unequivocal extraction of superconducting contributions. Prominent examples of such problems include the analysis of the Nernst effect [8,9] and of the optical conductivity [11,13], where a charge-stripe related signal might be mistaken for superconducting fluctuations [28-30], or linear magnetization and conductivity measurements [10,31], where the normal-state behavior is assumed to be linear in temperature, which is not necessarily the case. Several schemes to systematically subtract the presumed normal-state contribution have been devised, mainly based on the suppression of superconductivity with external magnetic fields [3,5,14]. However, so far only two experimental techniques can claim to be genuinely sensitive only to superconducting signals: nonlinear torque magnetization [6] and nonlinear conductivity [7].

A number of recent experimental investigations consistently point to a simple picture for both the normal state [15-18] and the superconducting emergence regime [3,4,6,7]. Measurements of transport properties, such as the dc resistivity [17], Hall angle [15] magnetoresistivity [16], and optical experiments [18], clearly show that the mobile charge carriers behave as a Fermi liquid, even in strongly underdoped compounds. The dual observations that the magnetoresistivity obeys Kohler scaling with a $1/\tau \propto T^2$ scattering rate [16] and that the optical scattering rate exhibits conventional scaling with temperature and frequency [18] are particularly clear-cut signatures of Fermi-liquid transport. Moreover, magnetization [6], high-frequency linear conductivity [3-5] and nonlinear response measurements [7] indicate that the superconducting emergence regime is limited to a rather narrow temperature range above $T_c$ and, importantly, that it can be described with a simple percolation model [7].

In the present work, we start from the fact that the normal state displays robust Fermi-liquid behavior in a rather wide temperature (and doping) range. We subtract its contribution to the planar resistivity with a reliability approaching the background-free techniques. With the inherent sensitivity of the dc conductivity to superconducting contributions, this enables us to obtain highly precise insight into the emergence of superconductivity. In particular, we performed measurements of the direct current (dc) conductivity for the cuprate $HgBa_2CuO_{4+\delta}$ (Hg1201) in the underdoped part of the phase diagram. Hg1201 may be viewed as a model compound due to its simple tetragonal structural symmetry, with one $CuO_2$ layer per formula unit, and the largest optimal $T_c$ (nearly 100 K) of all such single-layer compounds [32]. Further evidence for the model nature of Hg1201 comes from the observation of a tiny residual resistivity [17,33], of Shubnikov-de-Haas oscillations [34,35], and of a small density of vortex pinning centers [33], which has enabled the measurement of the triangular magnetic vortex lattice [36]. Below the characteristic temperature $T^{**}$ ($T^{**} < T^*$; $T^*$ is the pseudogap temperature), the planar resistivity of Hg1201 exhibits quadratic temperature dependence, $\rho \propto T^2$, the behavior characteristic of a Fermi liquid [17].

We studied two Hg1201 samples with $T_c \approx 80$ K (the estimated hole doping level is $p \approx 0.11$) that were prepared following established procedures [37,33]. This particular doping level was chosen because of a relatively wide temperature range between $T^{**}$ and $T_c$ in which pure quadratic-in-temperature resistivity is seen, while being reasonably far away from the doping level ($p \approx 0.09$) where weak short-range CDW correlations are most prominent in Hg1201 [26,27,38].

Figure 1 shows dc resistivity data for one of the two samples along with the three characteristic temperatures $T_c$, $T^{**}$ and $T^*$. The purely quadratic behavior seen below $T^{**}$ is in agreement with the Fermi-liquid character of the mobile holes [15,17]. The considerable difference between $T^{**}$ and $T_c$ provides for an extremely simple way to assess the superconducting paraconductivity contribution. In order to subtract the normal-state signal and obtain the purely superconducting contribution above $T_c$, we fit $\rho(T) = \rho_0 + a_2 T^2$ to the resistivity data in a temperature range from 100 K to $T^{**} \approx 150$ K, where $\rho_0$ is the small residual resistivity (the estimated residual resistivity ratio is approximately 120) and $a_2$ a constant. The resultant value of $a_2 = 9.8(1)$ n$\Omega$cm/K$^2$ is consistent with previous measurements on Hg1201 [17]. A narrowing of the fit range by 10-20 K does not change the result of our analysis, which demonstrates the robustness of the procedure. Furthermore, if a power law of the form $\rho(T) = aT^\alpha$ is fit in the same temperature range, the exponent is $\alpha = 1.98(2)$, and when the temperature range is varied by $\pm 20$ K, it stays within 5% of this value. The fidelity of the quadratic fit is very high (Fig. 1), which demonstrates that indeed in this temperature range the only contribution to the resistivity is the Fermi-liquid temperature dependence. We may therefore safely extrapolate the fit to $T_c$ in order to obtain the underlying normal-state contribution.

Inversion of the experimentally determined resistivity and subtraction of the extrapolated quadratic temperature dependence then gives the superconducting paraconductivity contribution, $\Delta\sigma_{dc}$, shown in Fig. 2. We present results for two samples (A and B), which were chosen from a larger batch of samples with $T_c \approx 80$ K due to their well-defined superconducting transitions – in Hg1201, the sample-contacting procedure often induces spurious doping of the sample surface [17], which can 'short out' the current path at temperatures above the nominal $T_c$ and artificially broaden the transition. Such samples are not considered here, although even they give similar results, except in a narrow (less than 1-2 K) temperature range above $T_c$. For samples A and B, we also performed magnetic susceptibility measurements using vibrating sample magnetometry (VSM) that show sharp transitions, with $T_c$ values that agree well with the resistive $T_c$. The zero-field-cool/field-cool susceptibility ratios approach one and are among the highest observed in the cuprates [33], demonstrating the very high quality of the samples (Fig. 1c and d). The excellent agreement of the paraconductivity results for these two distinct samples seen in Fig. 2, especially away from $T_c$, demonstrates the reproducibility of the experiment and the robustness of our result.

The superconducting response clearly exhibits exponential-like temperature dependence away from $T_c$, consistent with prior magnetization [6], nonlinear response [7], and microwave conductivity [5] results. We emphasize four crucial points: (i) the observed exponential dependence is qualitatively different from the underlying normal-state power-law behavior and hence a very robust result; (ii) the agreement with other experiments, some of which require no background subtraction [6,7], provides additional justification for the validity of our approach to subtract the Fermi-liquid normal state contribution; (iii) the signal-to-noise ratio of the present data is very high, which enables us to follow the paraconductivity over more than four orders of magnitude; (iv) both the exponential temperature dependence and the fact that the characteristic temperature is distinct from $T_c$ are incompatible with standard models of superconducting fluctuations, such as Ginzburg-Landau theory [39].

A simple superconducting percolation model with a compound-independent (and nearly doping-independent) underlying temperature scale $T_0$ (or energy scale $k_B T_0$) was recently shown to explain nonlinear response data [7]. The present dc paraconductivity result provides an ideal testing ground for this model, since the model is naturally formulated in terms of the dc conductivity. In particular, the model assumes that, above $T_c$, the material consists of patches that are normal and have a resistance $R_n$, and of patches that are superconducting and have a resistance $R_0$ (where we will take the limit $R_0 \rightarrow 0$) [7,40]. The fraction of superconducting patches, $P$, is temperature-dependent: at a critical fraction $P_\pi$ (corresponding to the critical temperature $T_\pi$), a sample-spanning superconducting cluster is formed, and hence percolates. In the limit of vanishingly small currents, $T_\pi$ equals $T_c$, but in any experiment, $T_c$ is shifted slightly below $T_\pi$ due to the required nonzero currents. The temperature-dependent superconducting fraction originates from an underlying distribution of superconducting gaps, and $P$ is hence directly obtained as the temperature integral of the distribution, as shown schematically in Fig. 3. For concreteness, we use the simplest (Gaussian) distribution with a full-width-at-half-maximum equal to $T_0$, consistent with previous work [7]. Other distributions, such as the gamma or the logistic distributions, were also tested, but resulted in no significant differences in the outcome of the calculation – slight discrepancies between calculations with different distributions only appear in the temperature range in which the signal is close to the noise level. This insensitivity to distribution shape is simply the result of the integration over the whole distribution, rendering the exact shape unimportant. The dc conductivity in dependence on temperature is now obtained using effective medium theory (EMT) [19], in a form derived specifically for site percolation problems [20]. While it is known that EMT becomes unreliable in the critical regime close to the percolation threshold [19] (in our case, about 1 K above $T_c$), we use it for simplicity and accuracy in the interesting higher-temperature regime away from $T_c$. The narrow critical regime is presumably not purely percolative anyway, with critical exponents modified by thermal effects [41]; in order to see a discrepancy between the data and the EMT calculation, a careful power-law analysis of the critical regime would need to be undertaken, with more closely-spaced measurements around $T_c$. The investigation of criticality is thus not within the scope of the present work.

In order to obtain the limit of zero $R_0$, we use different small values in the numerical calculation, until no significant changes in the output are seen (typically for $R_0$ on the order of $10^{-5} R_n$). We take $R_n$ to be constant in the temperature interval of interest – this is appropriate, because its relative change (due to the $T^2$ dependence) is about 25% over a 10 K interval, whereas the paraconductivity changes by a factor of about $10^2$ in the same interval. The calculated temperature dependence shown in Fig. 2 closely matches the experimental findings over the entire range of about four orders of magnitude in $\Delta\sigma$. As demonstrated in Fig. 2, the agreement between the data and the calculation can be further improved by adding a small offset in order to account for the crossover to the noise level.

Effectively, the percolation calculation of the paraconductivity only has one free parameter: the width $k_B T_0$ of the gap distribution. Other parameters that enter the calculation are constrained: $R_n$ is simply the normal-state resistivity, $T_\pi$ is slightly larger than $T_c$ (in the present calculation it was taken to be $T_c + 1$ K, but we note that our definition of $T_c$ as the lowest temperature with non-zero resistivity is somewhat arbitrary; different definitions, such as the midpoint of the transition measured by susceptibility, easily lead to a 1 K difference), and the critical concentration $P_\pi$ was taken to be 0.3, consistent with the prior nonlinear conductivity analysis [7]. The critical concentration $P_\pi$ is not arbitrary; it is determined by the details of the percolation model [42] – site or bond percolation, percolation with or without farther-neighbor corrections, etc. – and by the dimensionality of the percolation process. The model yields virtually the same temperature dependence for different values of $P_\pi$, with a corresponding change in $T_0$: a smaller $P_\pi$ implies a larger $T_0$, and vice versa. We therefore cannot distinguish among specific percolation scenarios, such as two-dimensional versus three-dimensional percolation. Prior comparison between linear and nonlinear response indicated that a three-dimensional site percolation model with $P_\pi \approx 0.3$ is appropriate [7], leading us to use the same value here. Remarkably, the value $T_0 = 26(1)$ K that yields the best agreement with the data in Fig. 2 is in excellent agreement with nonlinear conductivity and microwave linear response for a number of cuprate compounds and a range of doping levels, including Hg1201 [7].

The present work does not provide microscopic insight into the gap inhomogeneity and its origin, and in this respect the percolation model is phenomenological. Yet the model is highly consistent with experiments sensitive to real-space superconducting gap disorder, such as scanning tunneling microscopy [43,44], which have observed gap distributions with a width comparable to $k_B T_0$. It is furthermore consistent with NMR results that demonstrate a considerable distribution of local electric field gradients [45,46], and with X-ray experiments that find percolative structures in oxygen-doped $La_2CuO_{4+\delta}$ and $YBa_2Cu_3O_{6+\delta}$ [47,48].

In conclusion, for the simple-tetragonal cuprate Hg1201 the paraconductivity is a very sensitive probe of the emergence of superconductivity, and it is accurately described by the superconducting percolation scenario, with the same universal characteristic temperature scale observed for other observables [4-7]. We demonstrate that the superconducting contribution can be simply obtained upon assuming a Fermi-liquid normal state below the characteristic temperature $T^{**}$. This procedure is not possible for optimally-doped compounds, where $T^{**}$ becomes comparable to, or smaller than $T_c$ and the resistivity no longer exhibits quadratic temperature dependence [17]. It also is not possible for compounds such as the bismuth-based cuprates or twinned $YBa_2Cu_3O_{6+\delta}$, in which the underlying quadratic Fermi-liquid temperature dependence is masked due to disorder effects and/or low structural symmetry [15,16]. However, the clear confirmation of the superconducting percolation scenario in the present work implies that, fundamentally, both the normal-state carriers and superconducting emergence are rather conventional in the underdoped cuprates, once the underlying gap disorder is taken into account. Our result excludes the possibility of extended fluctuations usually associated with non-Fermi liquid models [8-10,49,50]. It also shows that it is difficult to observe the usual Ginzburg-Landau fluctuation regime in the conductivity, because inhomogeneity effects dominate – the percolation description holds down to temperatures very close to $T_c$. Along with magnetometry as well as linear and nonlinear conductivity data, the basic percolation model naturally explains other seemingly unconventional features such as the 'gap filling' seen in photoemission data [51], and thus provides a unifying understanding of superconducting pre-pairing in the cuprates [7]. The dc conductivity measurements presented here have put the scenario to a stringent quantitative test, and hence constitute a crucial, independent confirmation in a model cuprate system. The robustness of our result mandates a paradigm change in the field of cuprate superconductivity, namely that the itinerant carriers are well described by Fermi-liquid concepts, whereas the emergence of superconductivity is dominated by the gap inhomogeneity inherence to these lamellar oxides.

**Acknowledgments**

P.P. acknowledges funding by the Croatian Science Foundation Project No IP- 2016-06-7258. D.P. and M.P. acknowledge funding by the Croatian Science Foundation under grant no. IP-11-2013-2729. The work at the University of Minnesota was funded by the Department of Energy through the University of Minnesota Center for Quantum Materials, under DE-SC-0016371 and DE-SC-0006858. The work at the TU Wien was supported by FWF project P27980-N36.

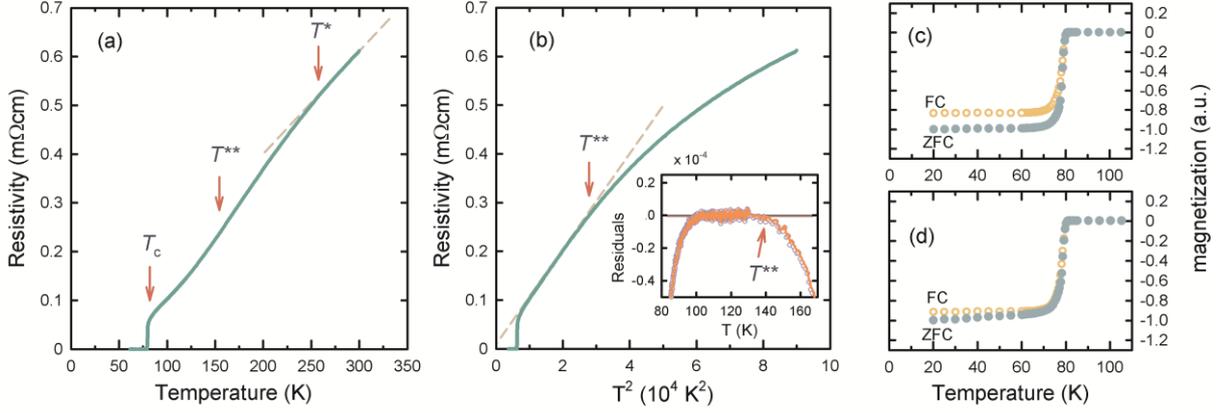

**FIG 1**. (a) Temperature dependence of the dc resistivity of a Hg1201 single crystal with characteristic temperatures $T_c \approx 80$ K (defined here to correspond to the lowest measurable non-zero resistivity), $T^{**} \approx 140$ K (defined as the deviation from low-temperature quadratic behavior), and $T^* \approx 260$ K (defined as the deviation from high-temperature linear behavior). (b) The dc resistivity, plotted versus the square of temperature and fit to $\rho = \rho_0 + aT^2$ between 100 K and 150 K (dashed line), demonstrating Fermi-liquid behavior below $T^{**}$ and a very small residual resistivity in the zero-temperature limit (see text). The inset shows the residuals obtained upon subtracting the fit result from the data for fits between 100 K and 150 K (line) and between 110 K and 130 K (symbols). The $T^2$ behavior prevails over a ~ 50 K range. (c) and (d) normalized VSM magnetization measurements of samples A and B, respectively, obtained with an external field of 15 Oe applied perpendicular to $CuO_2$ planes. Grey solid circles: zero-field-cooled (ZFC) data; orange open circles: field-cooled (FC) data. The measurements demonstrate well-defined superconducting transitions at $T_c \approx 80$ K and very low vortex pinning, indicative of high sample quality.

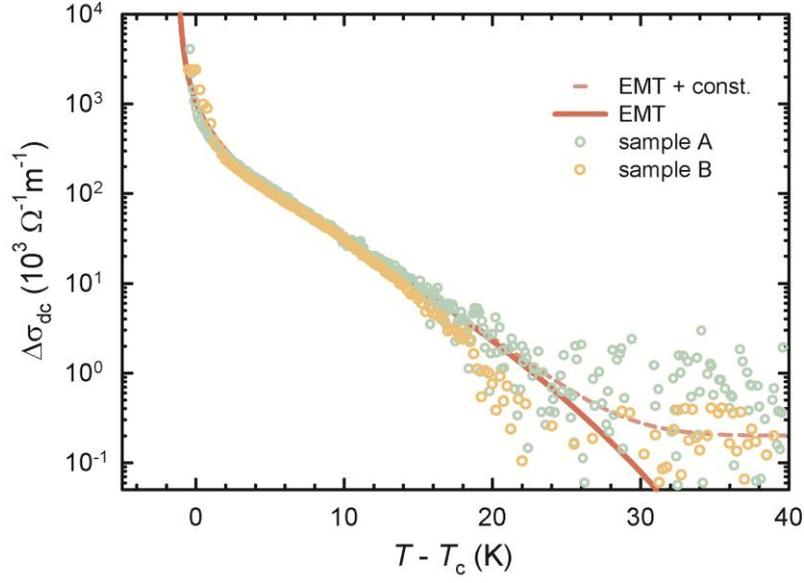

**FIG 2**. The dc paraconductivity for two underdoped Hg1201 samples (A and B) with $T_c \approx 80$ K, obtained by subtracting Fermi-liquid normal-state behavior from the measured resistivity. The very good agreement between the two data sets demonstrates a high level of reproducibility and robustness of the result. The paraconductivity exhibits strong exponential-like temperature dependence. The full line is the prediction of the superconducting percolation model obtained with effective medium theory. The dashed line includes a small heuristic constant offset and better captures the crossover to the noise level around $T \approx T_c + 25$ K (since, on a logarithmic scale, only the positive noise in $\Delta\sigma_{dc}$ is visible). $T_c$ is defined here as the lowest temperature at which a non-zero conductivity was measurable, whereas $T_\pi$ is the temperature at which the calculated conductivity diverges. $T_\pi$ is slightly larger than $T_c$ due to the nonzero current required to perform the experiment (see text).

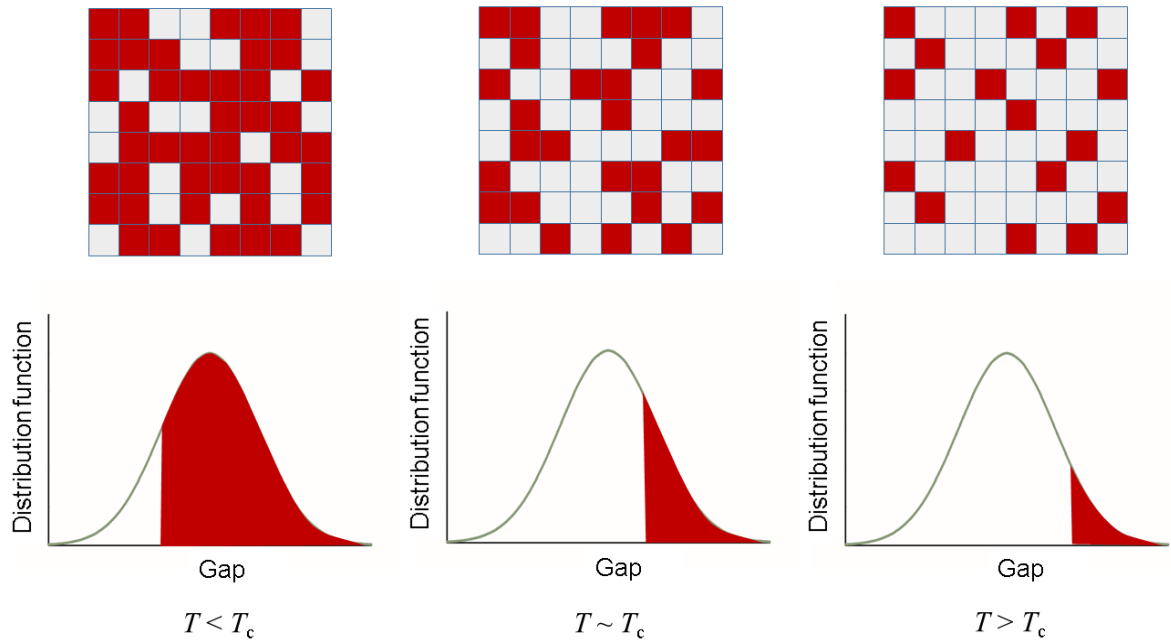

**FIG 3**. Schematic representation of the superconducting site percolation model, as a two-dimensional cross-section of the full three-dimensional model (upper row). Dark red patches are superconducting with vanishing resistance, whereas light grey patches have nonzero normal-state resistance. The fraction of superconducting patches is simply obtained by integrating the local gap distribution function taken to be a Gaussian for simplicity (lower row). Note that for a typical three-dimensional percolation model the critical fraction is approximately 0.3.